\documentclass[preprint,aps,amsmath,superscriptaddress,nofootinbib]{revtex4}

\usepackage{graphicx}
\usepackage{dcolumn}
\usepackage{bm}
\usepackage{textcomp}
\usepackage{threeparttable}
\usepackage{booktabs}
\usepackage{slashed}
\usepackage[dvips]{color}
\renewcommand{\TPTtagStyle}%
{\normalsize\textit}

\begin{document}

\preprint{INT-PUB-11-055}

\title{On the Role of Charmed Meson Loops in Charmonium Decays}
\author{Thomas Mehen and Di-Lun Yang}
\affiliation{Department of Physics, Duke University, Durham, NC 27708, USA}
\date{\today}

\begin{abstract}

We investigate the effect of intermediate charmed meson loops on the M1 radiative decays  $J/\psi \to \eta_c \gamma$ and $\psi'\rightarrow\eta^{(\prime)}_c\gamma$ as well as the isospin violating  hadronic decays $\psi'\rightarrow J/\psi \,\pi^0(\eta)$ using  heavy hadron chiral perturbation theory (HH$\chi$PT). The calculations include tree level as well as one loop diagrams and are compared to the latest data from CLEO and BES-III. Our fit constrains the couplings of 1S and 2S charmonium multiplets to charmed mesons, denoted $g_2$ and $g_2^\prime$, respectively. We find that there are two sets of solutions
for $g_2$ and $g_2^\prime$. One set, which agrees with previous values of the product $g_2 g_2^\prime$ extracted from analyses that consider only loop contributions to  $\psi'\rightarrow J/\psi \,\pi^0(\eta)$, can only fit data on radiative decays with fine-tuned cancellations between tree level diagrams and loops in that process. The other solution for $g_2$ and $g_2^\prime$  leads to couplings that are smaller by a factor of  2.3.  In this case tree level and loop contributions are of comparable size and the numerical values of the tree level contributions to radiative decays are consistent with estimates based on the   quark model as well as non-relativistic QCD (NRQCD). This result shows that  tree level HH$\chi$PT couplings are as important as the one loop graphs with  charmed mesons in these charmonium decays. The couplings $g_2$ and $g_2^\prime$ are also important for the calculations of the decays  of charmed meson bound states, such as the $X(3872)$, to conventional charmonia. 

\end{abstract}
\newpage

\maketitle

Many of the static properties and decays of charmonium states can be understood within a framework  in which these states are viewed as non-relativistic bound states of charm  and anticharm quarks. This includes the quark model~\cite{Eichten:1974af} as well as the modern QCD-based approach of non-relativistic QCD (NRQCD)~\cite{Bodwin:1994jh}, which allows for systematic treatment of charmonium properties as an expansion in $\alpha_s$ and $v_c$, where $v_c$ is the relative velocity of the charm-anticharm quarks. Despite many successes there remain specific transitions that are not well understood quantitatively. Examples of decays that are not completely understood are the hadronic decays $\psi^\prime \to J/\psi (\pi^0,\eta)$, and the radiative decays to $J/\psi \to \eta_c \gamma$ and $\psi^\prime \to \eta^{(\prime)}_c 
\gamma$. The hadronic decays violate isospin, in the case of a final state with $\pi^0$,  or $SU(3)$, when the final state is $\eta$.  As a consequence the ratio of this decay is sensitive to light quark masses~\cite{Ioffe:1980mx,Pham:1983mh}. The value of the light quark mass ratio  extracted from the measured decay rates~\cite{Bai:2004cg, :2008kb}, $m_u/m_d =0.4\pm 0.01$, differs significantly from the result extracted from meson masses in chiral perturbation theory, $m_u/m_d =0.56$~ \cite{Weinberg:1977hb, Gasser:1982ap}. For the radiative decays the experimentally measured rates differ from quark model expectations. For example, a non-relativistic quark model calculation of $J/\psi  \to \eta_c \gamma \, (\psi^\prime \to \eta_c \gamma)$ yields a prediction of $\approx 3$ ($\approx 0$) keV,~\footnote{The  decay rate  $\psi^\prime \to \eta_c\gamma$ vanishes in the non-relativistic quark model due to the vanishing overlap of the orbital wavefunctions of the $\psi^\prime$ and the $\eta_c$, and is no longer zero once relativistic corrections are taken into account. However, quark models that include relativistic corrections still
have trouble reproducing the correct rate for $\psi^\prime \to \eta \gamma$~\cite{PhysRevD.72.054026}.} whereas the experimental results are 1.57 $\pm$ 0.38  (0.97 $\pm$ 0.14) keV~\cite{Nakamura:2010zzi}. In Ref.~\cite{Brambilla:2005zw}, NRQCD is used to analyze the decay $J/\psi\to \eta_c\gamma$, and the authors show that  $O(v^2)$ corrections can lower the rate so that the theoretical prediction is consistent with data. However, no attempt has been made to understand radiative decays of $\psi^\prime$ in this framework. For reviews of these puzzles and others in charmonium physics, see Refs.~\cite{Voloshin:2007dx, PhysRevD.72.054026, Brambilla:2010cs}. 

Recently, Ref.~\cite{Guo:2009wr}  proposed that  the hadronic decays mentioned above are dominated by loop diagrams with virtual $D$ mesons. The decays are calculated using Heavy Hadron Chiral Perturbation Theory (HH$\chi$PT)~\cite{PhysRevD.45.R2188, PhysRevD.46.1148, Burdman:1992gh}, in which the charmonia are treated non-relativistically and  coupled to the $D$ mesons and Goldstone bosons in a manner consistent with heavy quark and chiral symmetries. In a non-relativistic theory the $D$ meson kinetic energy scales as $m_D v^2$ and momentum scales as $m_D v$, where $m_D$ is a $D$ meson mass and $v \approx 1/2$ is the typical velocity of the $D$ mesons in the loops. With this scaling, Ref.~\cite{Guo:2009wr} showed that the loop diagrams with $D$ mesons should be enhanced over tree level couplings by a factor of $1/v$.  The rates for $\psi^\prime \to J/\psi \pi^0$ and $\psi^\prime \to J/\psi \eta$ are sensitive to the product $g_2 g_2^\prime$, where the $J/\psi$ coupling to $D$ mesons is $g_2$  and the $\psi^\prime$ coupling to $D$ mesons is $g_2^\prime$. Ref.~\cite{Guo:2009wr}  found a value of $g_2 g_2^\prime$  consistent within errors with the two experimentally measured  rates. This resolves  the disagreement between the value of $m_u/m_d$ extracted from these decays and other extractions, since the prediction for the ratio of rates in terms of $m_u/m_d$ relied on the  rates being dominated by the tree level HH$\chi$PT coupling. The value of $g_2 g_2^\prime$ extracted by Ref.~\cite{Guo:2009wr}   is consistent with power counting estimates of $g_2$ and $g_2^\prime$, which are both expected to be $\sim (m_c v_c)^{-3/2}$ up to constants of order unity. Other hadronic and radiative charmonium decays are also analyzed within the same formalism in Refs.~\cite{Guo:2010ak,Guo:2010zk,Guo:2011dv}. 

The goal of this paper is to apply the same theory to the radiative decays mentioned above. One of our aims is to check whether the theory can also successfully resolve puzzles in radiative decays as one would hope. It is also important to check  that couplings extracted from the hadronic decays are  consistent with data on radiative decays.   An important aspect of our analysis is that unlike Refs.~\cite{Guo:2009wr,Guo:2010ak},  tree level counterterms are included in our calculations of  both hadronic and radiative  decays. Ref.~\cite{Guo:2010ak} argued for an additional factor in the loop graphs of $1/(4 \pi v_c^3)\approx 0.5-0.6$, for $v_c^2 \approx 0.25 -0.3$, which would  compensate the $1/v$ enhancement of the loops. This factor, and the fact that $v$  is not very small, support including both the loops and tree level interactions in the calculation, which we will do in this paper.
This can have an important impact of the extracted values of the couplings $g_2$ and $g_2^\prime$. Finally, an additional motivation for our analysis is that the extracted couplings are important for  the physics of the $X(3872)$ and other recently discovered charmonium bound states that have been interpreted as charmed meson molecules. If the $X(3872)$ is a charmed meson bound state, then the coupling $g_2 (g_2^\prime)$ is an important theoretical input for calculations of $X(3872) \to J/\psi (\psi^\prime)+ X $, so extraction of $g_2$ and $g_2^\prime$ is relevant to unconventional as well as conventional charmonia. For theoretical calculations of $X(3872)$ to conventional charmonia using effective field theory, see Refs.~\cite{Fleming:2008yn, Mehen:2011ds, Fleming:2011xa}.

Our main result is that in order to obtain a consistent fit to both radiative decays as well as the hadronic decays considered in Refs.~\cite{Guo:2009wr,Guo:2010ak}, counterterm contributions must be included and the values of $g_2$ and $g_2^\prime$ will then be smaller than estimated in an analysis containing only the loop diagrams by a factor of 2.3. This decreases the overall size of the loop amplitude by a factor of 5. It is not possible to get reasonable agreement with radiative decay data without including counterterms. Since NRQCD is the microscopic theory of  charmonia, and does not include loop effects from charmonia,  one is tempted to  identify the result of a calculation of the $J/\psi \to \eta_c \gamma$ amplitude in NRQCD with the tree level coupling in HH$\chi$PT. This is somewhat tenuous as the bare coupling in our theory has an infinite piece that  must cancel the linear divergence in the meson loop integrals.~\footnote{This linear divergence is absent in dimensional regularization.} Nevertheless, we regard it as satisfying that 
the size of the counterterms we extract in our fit with the smaller values of $g_2$ and $g_2^\prime$ are consistent within a factor of 2  with the quark model and NRQCD calculations of the radiative transitions.  For other extractions of the  couplings $g_2$ and $g_2^\prime$ in different theoretical frameworks, see, e.g., Refs.~\cite{PhysRevC.58.2994, PhysRevC.62.034903, PhysRevC.63.065201, PhysRevD.68.034002, Bracco:2004rx, Matheus:2002nq, PhysRevD.69.054023}. In Refs. \cite{Li:2007xr, Li:2011ss}, the charmed meson loop corrections to radiative $J/\psi$ and $\psi^\prime$  decays are studied in a version of  HH$\chi$PT with relativistic propagators and couplings, as well as form factors at the vertices that regulate ultraviolet divergences. The form factors introduce an additional parameter into the calculations. These authors did not attempt to simultaneously fit the hadronic decays but used values of  $g_2$ and $g_2^\prime$ consistent with those obtained in Refs.~\cite{Guo:2009wr,Guo:2010ak}. Their results are also consistent with the experimental data on the radiative decays.


The  effective HH$\chi$PT Lagrangian relevant to the hadronic decays is~\cite{Hu:2005gf, Fleming:2008yn, Guo:2010ak}
\begin{eqnarray}
\label{Lag}
\mathcal{L}&=&Tr[H^{\dagger}_{a}\left(i\partial_{0}+\frac{\nabla^{2}}{2m_{D}}\right)H_{a}]+\frac{\Delta}{4}Tr[H^{\dagger}_{a}\vec{\sigma}H_{a}\vec{\sigma}] -\frac{g}{2}Tr[H^{\dagger}_aH_b\vec{\sigma}\cdot\vec{u}_{ab}]
 \\
\nonumber
&+&i\frac{A}{4}\left(Tr[J'\sigma^iJ^{\dagger}]-Tr[J^{\dagger}\sigma^iJ']\right)\partial^i(\chi_-)_{aa} 
+ i\frac{g_{2}}{2}Tr[J^{\dagger}H_{a}\vec{\sigma}\cdot\overleftrightarrow{\partial}\bar{H}_a]+ H.c. \, .
\end{eqnarray}
Here  $H_{a}=V_{a}\cdot\vec{\sigma}+P_a$ and $\bar{H}_{a}=-\bar{V}_{a}\cdot\vec{\sigma}+\bar{P}_a$ are the charmed and anti-charmed meson multiplets with $V_a$ and $P_a$ denoting the vector and pseudoscalar charmed mesons, respectively, and  $J^{(\prime)}=\vec{\psi}^{(\prime)}\cdot\vec{\sigma}+\eta^{(\prime)}_c$ denotes the charmonium multiplets with $\vec{\psi}^{(\prime)}$ and $\eta_c^{(\prime)}$. The $\vec{\sigma}$ are the Pauli matrices, $a$ and $b$ denote flavor indices, and $A\overleftrightarrow{\partial}B=A(\vec{\partial}B)-(\vec{\partial}A)B$.  The first two terms in Eq.~(\ref{Lag}) are kinetic terms for the charmed mesons, $\Delta=m_{D^*}-m_{D}$ is the hyperfine splitting, and $m_D (m_{D^*})$ is the mass of pseudoscalar (vector) charmed meson.  The third term contains the interactions of $D$ mesons with the Goldstone boson fields which are contained in $u=exp(i\phi/\sqrt{2}F)$ where $\phi$ is a $3\times 3$ matrix of Goldstone boson fields and $F=92.4$ MeV is the pion decay constant. There are identical terms for the $\bar{D}$ mesons which are not explicitly shown.
The tree level couplings for $\psi'\rightarrow J/\psi\pi^0(\eta)$ come from the term with coupling constant $A$. The factor $\chi_-$ is defined by
 $\chi_-=u^{\dagger}\chi u^{\dagger}-u\chi^{\dagger} u$, where  $\chi=2B_0\cdot diag(m_u,m_d,m_s)$,  $m_u$ $m_d$ and $m_s$ are the light  quark masses and $B_0=|\langle 0|\bar{q}q|0\rangle|$. 
 Finally, the 1S charmonia couple to the $D$ mesons via the last term with coupling $g_2$. The same term, with $J$ and $g_2$ replaced with  $J^\prime$ and $g^\prime_2$, couples the $2S$ charmonia to charmed mesons.
 
The tree level decay amplitudes are~\cite{Guo:2010ak},
\begin{eqnarray}\label{amps}
\nonumber
i\mathcal{M}(\psi'\rightarrow J/\psi\pi^0)&=&i4A\epsilon_{ijk}q_i\epsilon^{\psi'}_j\epsilon^{J/\psi}_k B_{du}\\
i\mathcal{M}(\psi'\rightarrow J/\psi\eta)&=&i(8/\sqrt{3})A\epsilon_{ijk}q_i\epsilon^{\psi'}_j\epsilon^{J/\psi}_k B_{sl},
\end{eqnarray}
where $B_{du}=\frac{B_0}{F}(m_d-m_u)$ and $B_{sl}=\frac{B_0}{F}(m_s-\frac{m_u+m_d}{2})$. To leading order in the chiral expansion, these factors may be expressed in terms of light meson masses: $B_{du}=(m_{K^0}^2-m_{K^+}^2+m_{\pi^+}^2-m_{\pi^0})/F$ and $B_{sl}=(3/4)(m_{\eta}^2-m_{\pi_0}^2)/F$. The $\pi^0-\eta$ mixing must also be included, and the mixing angle is
\begin{eqnarray}
\epsilon_{\pi^0\eta}=\frac{1}{\sqrt{3}}\frac{m_{K^0}^2-m_{K^+}^2+m_{\pi^+}^2-m_{\pi^0}}{m_{\eta}^2-m_{\pi_0}^2}.
\end{eqnarray}
When this mixing is included the first matrix element in Eq~(\ref{amps}) is multiplied by 3/2. The loop diagrams contributing to the decay have been evaluated in Refs.~\cite{Guo:2009wr,Guo:2010ak}. Since the decay to $\pi^0(\eta)$ vanishes in the isospin ($SU(3)$) limit, the diagrams cancel in the sum over $D^0, D^+$, and $D_s^+$ appearing in the loop in the limit that all these mesons are degenerate. Mass differences between the mesons render the cancellation incomplete and are responsible for the finite contribution.

For electromagnetic decays, we need to add couplings to the magnetic field and gauge the interactions in Eq.~(\ref{Lag}). The tree level coupling of the charmonia to the magnetic fields is given by~\cite{DeFazio:2008xq,Casalbuoni:1992yd}
\begin{eqnarray}
\frac{\rho}{2}Tr[J\vec{B}\cdot\vec{\sigma}J^{\dagger}] +\frac{\rho^\prime}{2}(Tr[J^\prime\vec{B}\cdot\vec{\sigma}J^{\dagger}] +H.c.)
+\frac{\rho^{\prime \prime}}{2}Tr[J^\prime\vec{B}\cdot\vec{\sigma}J^{\prime \,\dagger}] \, ,
\end{eqnarray}
where $\vec{B}$ is the magnetic field. Due to the presence of Pauli matrices these terms break  heavy quark spin symmetry. The first term is responsible for the decay $J/\psi \to \eta_c \gamma$, the second for $\psi^\prime \to \eta_c \gamma$, and the third for  $\psi^\prime \to \eta^\prime_c \gamma$.
For the loop corrections to the radiative decays, we must also include the coupling of the  charmed mesons to the magnetic field, which is given by \cite{Amundson:1992yp,Hu:2005gf}
\begin{eqnarray}\label{mag}
\frac{e\beta}{2}Tr[H^{\dagger}_{a}H_{b}\vec{\sigma}\cdot\vec{B}Q_{ab}]
+\frac{e}{2m_{c}}Q'Tr[H_{a}^{\dagger}\vec{\sigma}\cdot\vec{B}H_{a}],
\end{eqnarray}
where $Q_{ab}=diag(2/3,-1/3,-1/3)$, $Q'=2/3$, and $m_c$ is the mass of charm quark. These terms are responsible for the decays $D^{*} \to D \gamma$.
Including leading as well as $\Lambda_{QCD}/m_c$ suppressed terms is crucial for reproducing observed $D^* \to D\gamma$ rates~\cite{Amundson:1992yp}.
 Ref.~\cite{Hu:2005gf} finds that a good fit to the experimental rates 
is obtained for the values $m_c = 1.5$ GeV and $\beta = 3.0 \, \rm{GeV}^{-1}$.

\begin{figure}
{\includegraphics[width=1.0\textwidth]{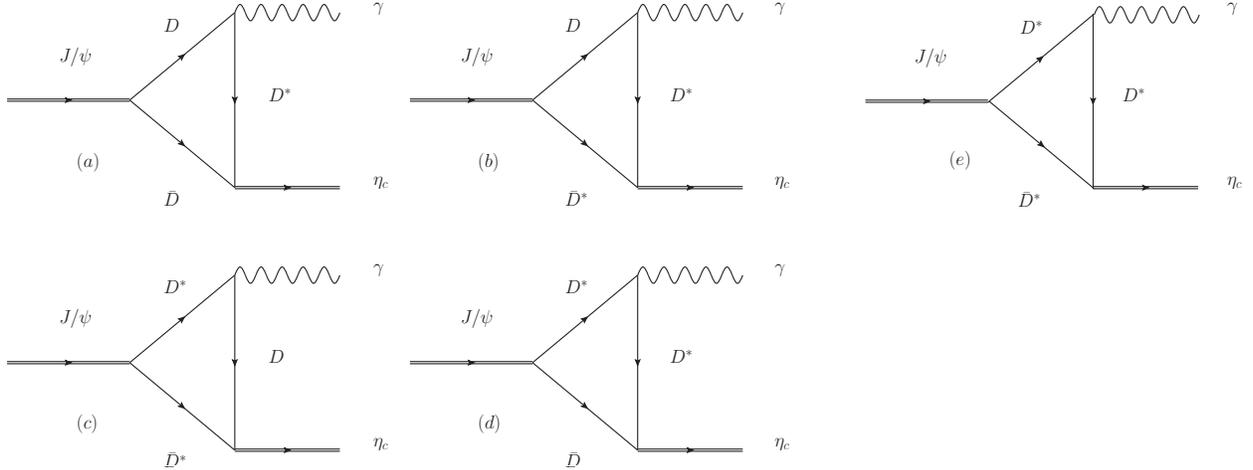}}
\caption{Triangle diagrams with intermediate charmed meson loops. The charmed meson couplings to photon come from Eq.~(\ref{mag}).}
\label{triangle}
\vspace{3mm}
\end{figure}

\begin{figure}
{\includegraphics[width=0.8\textwidth]{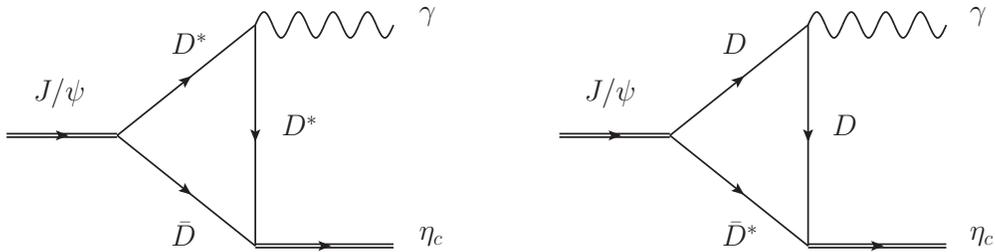}}
\caption{Triangle diagrams with intermediate charmed meson loops. The charmed meson couplings to photon come from gauging the kinetic terms in Eq.~(\ref{Lag}).}
\label{gauge}
\vspace{3mm}
\end{figure}

\begin{figure}
\hspace{-0.5in}{\includegraphics[width=0.6\textwidth]{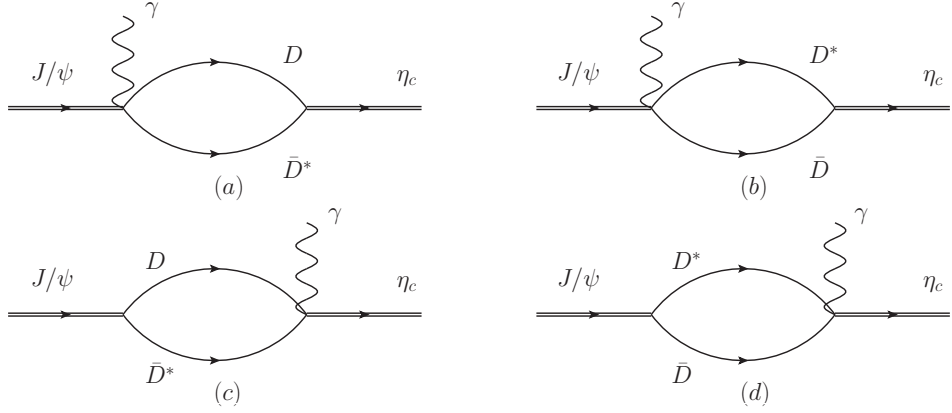}}
\caption{Contact diagrams with intermediate charmed meson loops,}
\label{contact}
\vspace{3mm}
\end{figure}

These couplings enter the radiative decays of charmonia through the triangle diagrams shown in
Fig.~\ref{triangle}.
There are also interactions that arise from gauging the derivatives in Eq.~(\ref{Lag}). Gauging the derivatives in the kinetic term for the $D$ mesons leads to couplings to the photon which contribute to the radiative decays via triangle loop diagrams shown in Fig.~\ref{gauge}.
Gauging the coupling $g_2^{(\prime)}$ leads to a contact interaction that directly couples charmonia, heavy mesons and the photon field, which is given by
\begin{eqnarray}\label{contactint}
- \,eg_2 Tr[J^{\dagger}H_{a}\vec{\sigma}\cdot\vec{A}\bar{H}_a]+H.c.
\end{eqnarray}
where $a =2$ or $3$ only, i.e., only charged and strange $D$ mesons appear in the interaction term of Eq.~(\ref{contactint}). The loop diagrams with contact interactions are shown in Fig.~\ref{contact}. 

The tree level amplitude for the  $J/\psi\rightarrow \eta_c \gamma$ decay, for example, is 
\begin{eqnarray}
i\mathcal{M}_0 = \rho\, \epsilon_{ijk} \,q_{i}\epsilon^{\gamma}_{j}\epsilon^{J/\psi}_{k}\, ,
\end{eqnarray}
where $q$ denotes the momentum of photon, and  $\epsilon^{\gamma}$ and $\epsilon^{J/\psi}$ are the polarization vectors of the photon and $J/\psi$, respectively.
The corresponding  decay rate is
\begin{eqnarray}\label{decay}
\Gamma[J/\psi\rightarrow\eta_c \gamma]&=&\frac{1}{8\pi}|\sqrt{m_{J/\psi}m_{\eta_c}} \mathcal{M}_0|^2\frac{|\vec{q}\,|}{m_{J/\psi}^2}, \nonumber \\
&=& \frac{\rho^2}{12 \pi} \frac{m_{\eta_c}}{m_{J/\psi}} |\vec{q}\,|^3 \, .
\end{eqnarray} 
Here the factor $\sqrt{m_{J/\psi}m_{\eta_c}}$ comes from the normalization of nonrelatvistic fields in HH$\chi$PT. In the non-relativistic quark model, 
$\rho  =  2 e e_c/m_c \approx 0.3 \,{\rm GeV}^{-1}$, and the predicted decay rate is about a factor of 2 too large. The contribution to the amplitude from 
meson loops is also proportional $\epsilon_{ijk}q_{i}\epsilon^{\gamma}_{j}\epsilon^{J/\psi}_{k}$ so the loops give an additive shift to the $\rho$ term for each decay.
We evaluate the loops in pure dimensional regularization so linear divergences do not appear and the corrections from all loops are finite.
The full rate is Eq.~(\ref{decay}) with $\mathcal{M}_0$ replaced by $\mathcal{M}_{full}$, where the  $\mathcal{M}_{full}$ includes both the tree level interaction and the contributions from neutral, charged, and strange meson loops.    The explicit expression for $\mathcal{M}_{full}$ can be found in the Appendix. 

Before proceeding to our fits to the data, we will briefly discuss the power counting for the diagrams we have shown.  As stated earlier, 
for non-relativistic $D$ mesons one takes $E\sim m_D v^2$, $p \sim m_D v$, so the propagators scale as $(m_D v^2)^{-1}$ and  the loop integration measure is $m_D^4 v^5$. The vertices coupling the charmonium to $D$ mesons carry a factor of $p \sim m_D v$. To estimate $v^2$ one may take the difference between an external and two internal mesons, so $m_D v^2 = |m_{\rm charmonium}- m_{D\,{\rm pair}}|$ where $m_{\rm charmonium}$ is the mass of one of the external charmonia and $m_{D\, {\rm pair}}$ is the mass of two $D$ mesons in the loop. This leads to an estimate ranging from $v^2 = 0.09$ ($m_{\rm charmonium}= m_{\psi^\prime}$ and $m_{D \,{\rm pair}} =m_{D^0} + m_{D^{*0}})$ to $v^2 = 0.5$ ($m_{\rm charmonium} = m_{J/\psi}$ and $m_{D\, {\rm pair}}= 2 m_{D^{*0}}$). Naively with this counting the triangle diagrams scale as $(m_D^4 v^5) (m_D v^2)^{-3} (m_D v)^2 q = m_D^3 v q$. The first factor comes from the loop integration factor, the second from the propagators,  the third  factor from the derivative couplings of charmonium to $D$ mesons, and the factor of $q$ is the photon or pion momentum which comes from the coupling of these particles to $D$ mesons.  The diagrams with the contact interaction  scale as $(m_D^4 v^5) (m_D v^2)^{-2} q = m_D^3 v q$ which is the same as the triangle graph. This is because there is no derivative in the contact interaction with the photon, the derivative in the charmonium $D$ meson coupling must turn into the factor of $q$ required by gauge invariance or chiral symmetry and there are only two propagators. The factor $q$ is common to all diagrams including the tree level diagrams. Factors of $m_D$ are compensated by other dimensionful couplings so we  will focus only on counting powers of $v$ from here on. So the triangle graphs and graphs with the contact interaction are $v$ suppressed relative to the tree level interactions. However, these hadronic decays violate either isospin or $SU(3)$ and the radiative decays violate  heavy quark symmetry so there are cancellations between graphs due to heavy meson mass differences that  are missed by this power counting. One can formally modify the power counting counting in the following way. The inverse propagator for a non-relativistic meson can be written as $E- \frac{p^2}{2 m_D} + b + \delta$, where the residual mass term in the propagator has been split into a term $b$ which is common to all $D$ meson states and a term $\delta$ contains $SU(3)$ breaking and hyperfine splittings that are different for different $D$ mesons. Expand the $D$ meson propagators as 
\begin{eqnarray}
\frac{1}{E - \frac{p^2}{2 m_D} +b +\delta}= \frac{1}{E -  \frac{p^2}{2 m_D} +b} - \frac{\delta}{\left(E-  \frac{p^2}{2 m_D} + b\right)^2} + ... \, . \nonumber
\end{eqnarray}
The graph in which all propagators contribute only the first term is zero by symmetry. In order to get a non-vanishing result at least one propagator in the graph must give a contribution from  the second term, then the power counting says the graph is enhanced by a factor of $\delta/m_D v^2$, which makes the graph $1/v$ enhanced rather than $v$ suppressed relative to the tree level diagrams~\cite{Guo:2009wr,Guo:2010ak}. Since $v$ is not very small this could be compensated by other numerical factors. In practice it is easier to simply calculate the graphs with the unexpanded propagators but expanding the propagator makes it clear that after summing over all graphs one gets a $1/v$ enhancement. In this paper, we will take the viewpoint that the leading one loop diagrams are of roughly the same size as the tree level contributions and include both in the decays, then try to simultaneously fit the radiative and hadronic decays mentioned above.

A separate question is whether higher order chiral corrections are under control. Certainly some chiral corrections are suppressed  as argued for different charmonium radiative decays in Ref.~\cite{Guo:2011dv}. But in a subgraph with a ladders of single pion exchanges between a pair of $D$ mesons, non-relativistic power counting shows that the ladder with $n+1$ single pion exchanges is suppressed relative to one with $n$ pion exchanges by a factor  $g^2 m_D p/(8 \pi F^2) =  p/(320 MeV)$ \cite{Kaplan:1998we} where $p$ is the relative momentum of the $D$ mesons. This would require $p = m_D v$ with $v < 0.08$ to be less than $1/2$. In some channels a resummation of single pion exchanges may be needed to do accurate calculations. Such a resummation is beyond the scope of this paper. Here we are simply interested in the impact that including the tree level interactions and simultaneously fitting the radiative and hadronic decays has on the values of $g_2$ and $g_2^\prime$ and therefore the size of $D$ meson loop contributions to charmonium decays.

To constrain the parameters $g_2$ and $g_2^\prime$ , we determine the parameter $A$ and the product $g_2 g_2^\prime$ from the measured rates for $\Gamma[\psi^\prime \to J/\psi \pi^0]$ and $\Gamma[\psi^\prime \to J/\psi \eta]$. Because the predictions for the decay rates are quadratic in $g_2 g_2^\prime$, this does not completely determine $g_2 g_2^\prime$, but yields two possible solutions.  Then we fix the relative size of the two couplings using the relation $g_2 = g_2^\prime \sqrt{m_\psi^\prime/m_{J/\psi}}$, which follows if the dimensionless coupling of the $J/\psi$ and $\psi^\prime$ to $D$ mesons is the same~\cite{Guo:2009wr,Guo:2010ak}. 
 Once $g_2$ and $g_2^\prime$ are determined this way from the hadronic decays, the only parameters remaining in the  radiative decays are $\rho$, $\rho^\prime$, and $\rho^{\prime\prime}$, which can be determined from the three decay rates $\Gamma[J/\psi \to \eta_c \gamma]$, $\Gamma[\psi^\prime \to \eta_c \gamma]$, and  $\Gamma[\psi^\prime \to \eta_c^\prime \gamma]$.  
\begin{table}[t]
\begin{tabular}{|l|c|c|c|c|c|}
\hline
$A$  $(GeV^{-2})$ & $g_2$  $(GeV^{-3/2})$& $g'_2$ $(GeV^{-3/2})$& $\rho$ $(GeV^{-1})$& $\rho''$ $(GeV^{-1})$& $\rho'$ $(GeV^{-1})$\\
\hline
$-0.00636^{+4\times 10^{-5}}_{-4\times 10^{-5}}$ &$1.37^{+0.02}_{-0.02}$ &$1.26^{+0.02}_{-0.02}$ &$1.86^{+0.09}_{-0.09}$ &$0.83^{+0.12}_{-0.12}$ & $1.56^{+0.05}_{-0.05}$ \\
\hline
$-0.00636^{+4\times 10^{-5}}_{-4\times 10^{-5}}$ &$1.37^{+0.02}_{-0.02}$ &$1.26^{+0.02}_{-0.02}$ &$2.27^{+0.09}_{-0.09}$ &$1.26^{+0.12}_{-0.12}$ & $1.59^{+0.05}_{-0.05}$ \\
\hline
$0.0257^{+0.0007}_{-0.0007}$ &$0.599^{+0.023}_{-0.023}$& $0.549^{+0.021}_{-0.021}$ & $0.191^{+0.054}_{-0.054}$ & $-0.0192^{+0.105}_{-0.105}$&$0.287^{+0.024}_{-0.024}$ \\
\hline
$0.0257^{+0.0007}_{-0.0007}$ &$0.599^{+0.023}_{-0.023}$& $0.549^{+0.021}_{-0.021}$ & $0.598^{+0.054}_{-0.054}$ & $0.415^{+0.105}_{-0.105}$&$0.313^{+0.024}_{-0.024}$ \\
\hline
 & &  & $0.275$ & $0.263$ &$0.0417$ \\
\hline
\end{tabular}
\caption{The numerical results for fitting parameters to hadronic and radiative charmonium decays. The fit is explained in the text. The quark model~\cite{PhysRevD.72.054026} predictions for the parameters $\rho$, $\rho^\prime$ and
$\rho^{\prime \prime}$ are shown in the bottom line. Errors are due to experimental uncertainties only.}
\vspace{0.25 in}
\end{table}
 The results of determining $A$, $g_2$, and $g_2^\prime$ are shown in the first three columns of Table 1. One possible fit to the hadronic decays yields $A = -6.36 \, 10^{-3}\, {\rm GeV}^{-2}$ and $g_2 g_2^{\prime} = 1.73\, {\rm GeV}^{-3}$. This is a very small value of $A$, almost two orders of magnitude smaller than the estimate $A\sim 1/(2 m_c^2)$ in Ref.~\cite{Guo:2010ak}. This fit yields a value of $g_2 g_2^\prime$ similar to that of Refs.~\cite{Guo:2009wr,Guo:2010ak}.~\footnote{In this case, we get a value of $g_2 g_2^\prime$ that is a factor of two smaller than Refs.~\cite{Guo:2009wr,Guo:2010ak} because  our calculations of the loop amplitudes for $\psi^\prime \to J/\psi  \pi^0 (\eta)$ disagree with the analytic results of Refs.~\cite{Guo:2009wr,Guo:2010ak} by an overall factor of two. This is because we include graphs in which the $\pi^0$ or $\eta$ couples to the $\bar{D}^{(*)}$ mesons, instead of the $D^{(*)}$ mesons, that are omitted in Refs.~\cite{Guo:2009wr,Guo:2010ak}.}
 The second  possible fit is $A = 2.57\, 10^{-2} \, {\rm GeV}^{-2}$ and $g_2 g_2^\prime = 0.329\, {\rm GeV}^{-3}$, a value 5.3 times smaller than the first fit. The value of $A$ is closer to the estimate of $1/(2m_c^2)$, but still a factor of 10 smaller. The results of fitting the parameters $\rho$, $\rho^\prime$, and $\rho^{\prime \prime}$, are shown in the last three columns of Table 1. For each choice of $A$, $g_2$, and $g_2^\prime$,  there are two possible solutions for $\rho$, $\rho^\prime$, or $\rho^{\prime\prime}$, for a total of four possible solutions. The values of $\rho$ and $\rho^{\prime \prime}$ are much closer to the quark model predictions 
  (shown in the bottom row of Table 1) in the fit with a smaller value of $g_2 g_2^\prime$. The extracted value of $\rho^\prime$ does not come close to the quark model prediction of Ref.~\cite{PhysRevD.72.054026}, but this model does not give a good prediction for the rate $\psi^\prime \to \eta_c \gamma$.
  In the fits with the larger value of $g_2 g_2^\prime$ the extracted values of $\rho$, $\rho^\prime$, and $\rho^{\prime\prime}$ are much larger. This indicates that for these choices of parameters fine tuned cancellations between the tree level and loop diagrams are required to fit the data. This can also be clearly seen in Table II, where we give the loop contribution to the decay for each fit. For the first fit with $g_2 g_2^{\prime} = 1.73\, {\rm GeV}^{-3}$, $\Gamma[J/\psi \to \eta_c \gamma]$ is over predicted by a factor of 100 and  $\Gamma[\psi^\prime \to \eta^\prime_c \gamma]$ is over predicted by a factor of 20  without the counterterm  contribution. Thus, in order to fit these decays, fine-tuned cancellations between loop and tree level contributions must occur. Though the loop contributions by themselves do not do a good job of producing the radiative decay rates for the smaller value of $g_2 g_2^\prime$, the discrepancy is not nearly as large. 

For the decay $\psi^\prime \to \eta_c \gamma$ there are rather severely fine tuned cancellations between loop diagrams and tree level contributions for both fits. 
The photon energies in the decays $J/\psi \to \eta_c \gamma$, $\psi^\prime \to \eta_c \gamma$, and $\psi^\prime \to \eta_c^\prime \gamma$ are 114 MeV, 638 MeV, and 49 MeV, respectively. The photon energy in the second decay may be too large for either the quark model or  low energy effective theory to be accurate.  As an alternative approach, one can simply try to extract $g_2$ and $g_2^\prime$ independently from the radiative decays 
$J/\psi \to \eta_c \gamma$ and $\psi^\prime \to \eta_c^\prime \gamma$, using the quark model~\cite{PhysRevD.72.054026} to estimate  the parameters $\rho$ and $\rho^{\prime \prime}$. Since the decay $J/\psi \to \eta_c \gamma$ is quadratic in $g_2^2$  and the decay $\psi^\prime \to \eta_c^\prime \gamma$ is quadratic in $g_2^{\prime \, 2}$, there are two possible solutions for each parameter. 
We find  $g_2=0.255^{+0.042}_{-0.042}$ GeV $^{-3/2}$ or $0.659^{+0.016}_{-0.016}$ GeV$^{-3/2}$ and $g'_2=0.264^{+0.260}_{-0.260}$ GeV$^{-3/2}$ or $0.855^{+0.080}_{-0.080}$ GeV$^{-3/2}$. 
Note that the value of $g_2 g_2^\prime$ obtained this way is also smaller than the value obtained in the first fit to the combined hadronic and radiative decays.
Also the ratio $g_2^\prime/g_2$ obtained using the the smaller two central values is $1.04$ while using the larger two central values the ratio is $1.3$. Both of these are a little larger than one expects from the
hypothesis $g_2^\prime/g_2 = \sqrt{m_{J/\psi}/m_{\psi^\prime}} = 0.92$.

\begin{table}[t]
\begin{tabular}{|l|c|c|c|c|}
\hline
 &Fit 1  & Fit 2 & QM & PDG, BES III, and CLEO\\
\hline
$\Gamma[J/\psi\rightarrow \eta_c \gamma]_{loop}$ &$163^{+10}_{-9}$ keV&$5.96^{+0.97}_{-0.86}$ keV&2.9 keV &1.58$\pm$ 0.37 keV\cite{Nakamura:2010zzi}\\
\hline
$\Gamma[\psi'\rightarrow \eta'_c \gamma]_{loop}$ &$3.30^{+0.21}_{-0.20}$ keV&$0.119^{+0.019}_{-0.017}$ keV&0.21 keV &0.143$\pm$ 0.027$\pm$0.092 keV\cite{Xmisc}\\
\hline
$\Gamma[\psi'\rightarrow \eta_c \gamma]_{loop}$ &$16.34 ^{+0.98}_{-0.93}$ MeV& $597^{+97}_{-87}$ keV&9.7 keV &0.97$\pm$ 0.14 keV\cite{:2008fb}\\
\hline
\end{tabular}
\caption{The decay rates contribution from the loops alone for the two solutions for $g_2 g_2'$ is compared with the results in the quark model~\cite{PhysRevD.72.054026} and experimental data.
Fit 1 corresponds to $g_2 g_2^\prime =1.73^{+0.05}_{-0.05}$ (GeV$^{-3}$), Fit 2 corresponds to $g_2 g_2^\prime=0.329^{+0.025}_{-0.025}$ (GeV$^{-3}$).}
\end{table}

In summary, we have computed the decay rates for  $J/\psi \to \eta_c \gamma$ and $\psi'\rightarrow \eta^{(\prime)}_c \gamma$ including tree level and one loop diagrams with charmed mesons in HH$\chi$PT. We combined our results with the decay rates for $\psi'\rightarrow J/\psi\pi^0(\eta)$ found in Refs.~\cite{Guo:2009wr,Guo:2010ak}, used the relationship $g_2^\prime = g_2\sqrt{m_{J/\psi}/m_{\psi^\prime}}$, and fit the five remaining coupling constants  simultaneously. 
Including tree level couplings is essential for simultaneously reproducing  all the decay rates. A smaller value of $g_2 g_2^\prime = 0.33 \, {\rm GeV}^{-3}$ is required to avoid large cancellations between tree level and charmed  meson loop contributions to the radiative decay. The tree level couplings $\rho$ and $\rho^{\prime \prime}$ in this fit are consistent (to within a factor of 2) with expectations based on the quark model and NRQCD.

\acknowledgements
This work was supported in part by the  Director, Office of Science,  Office of Nuclear Physics, of the U.S. Department of Energy under grant numbers  
DE-FG02-05ER41368 (T.M.)  and DE-FG02-05ER41367 (D.Y.). We thank the Department of Energy's Institute for Nuclear Theory at the University of Washington for its hospitality  during the completion of this work.

\section*{APPENDIX}
In this Appendix, we calculate the loop diagrams for the M1 radiative decays. We present the calculation of $J/\psi \to \eta_c \gamma$, the calculations of the  decays $\psi^\prime \to \eta_c \gamma$ 
($\psi^\prime \to \eta^\prime_c \gamma$) are obtained by replacing $m_{J/\psi}$ with $m_{\psi^\prime}$ and $g_2^2$ with $g_2 g_2^\prime$ ($g_2^{\prime\, 2}$).

 The triangle loop diagrams in Fig.~\ref{triangle} have a similar form as the triangle loop diagrams in the hadronic decays, therefore 
 the  notation used here will be the almost the same as that of Ref.~\cite{Guo:2010ak}. We refer the reader to that paper for explicit expressions for the integrals. 
The amplitude from  Fig.~\ref{triangle}(a) is
\begin{eqnarray}
\nonumber
i\mathcal{M}_{1a}&=&-2ig_{2}^2 \lambda_{1(3)}
\int\frac{d^{4}l}{(2\pi)^4}\frac{\epsilon_{ijk}q_{i}\epsilon^{\gamma}_{j}(2l-q)_{k}\vec{\epsilon}^{\,J/\psi}\cdot\vec{l}}{8(l_{0}-\frac{\vec{l}^2}{2m_D}+i\epsilon)(l_{0}+\frac{\vec{l}^2}{2m_D}+b_{DD}-i\epsilon)
(l_{0}-q_{0}-\frac{(\vec{l}-\vec{q})^2}{2m_{D^{*}}}-\Delta+i\epsilon)}\\
&=&4g_2^2\lambda_{1(3)}\epsilon_{ijk}q_{i}\epsilon^{\gamma}_{j}\epsilon^{J/\psi}_{k}|\vec{q}\,|^2I^{(2)}_1(q,m_D,m_D,m_{D^*})\, ,
\end{eqnarray}
where $b_{DD} = 2 m_D - m_{J/\psi}$, $\lambda_1=\frac{2}{3}(e\beta+\frac{e}{m_c})$ is relevant for loops with neutral  $D$ mesons and 
 $\lambda_3=-\frac{1}{3}(e\beta-\frac{2e}{m_c})$ is relevant for loops with charged and strange $D$ mesons.
  Here $I^{(2)}_1(q,m_1,m_2,m_3)$ only differs from the function defined in Ref.~\cite{Guo:2010ak} by omitting a factor of $m_1m_2m_3$ from  the denominator.
  Fig.~\ref{triangle}(b) contributes
  \begin{eqnarray}
i\mathcal{M}_{1b} &=& 2g_2^2\lambda_{1(3)}\epsilon_{ijk} q_{i}\epsilon^{\gamma}_{j}\epsilon^{J/\psi}_{k}|\vec{q}\,|^2  \\
&\times& (2I^{(2)}_0(q,m_D,m_{D^*},m_{D^*})+4I^{(2)}_1(q,m_D,m_{D^*},m_{D^*})-I^{(1)}(q,m_D,m_{D^*},m_{D^*}))\, , \nonumber
\end{eqnarray}
where the functions $I^{(2)}_0(q,m_1,m_2,m_3)$ and $I_1(q,m_1,m_2,m_3)$ are again the same as functions in Ref.~\cite{Guo:2010ak} up to a factor of $m_1m_2m_3$.
Fig.~\ref{triangle}(c) contributes
\begin{eqnarray}
i\mathcal{M}_{1c} &=& 2g_2^2\lambda_{1(3)}\epsilon_{ijk} q_{i}\epsilon^{\gamma}_{j}\epsilon^{J/\psi}_{k}|\vec{q}\,|^2  \\
&\times&(2I^{(2)}_0(q,m_{D^*},m_{D^*},m_D)+6I^{(2)}_1(q,m_{D^*},m_{D^*},m_D)-I^{(1)}(q,m_{D^*},m_{D^*},m_D))\, . \nonumber
\end{eqnarray}
Fig.~\ref{triangle}(d) contributes
\begin{eqnarray}
i\mathcal{M}_{1d} &=& 2g_2^2\lambda_{2(4)}\epsilon_{ijk} q_{i}\epsilon^{\gamma}_{j}\epsilon^{J/\psi}_{k}|\vec{q}\,|^2  \\
&\times&(2I^{(2)}_0(q,m_{D^*},m_D,m_{D^*})+4I^{(2)}_1(q,m_{D^*},m_D,m_{D^*})-I^{(1)}(q,m_{D^*},m_D,m_{D^*})) \, , \nonumber
\end{eqnarray}
where $\lambda_2=-\frac{2}{3}(e\beta-\frac{e}{m_c})$ is relevant for loops with neutral  $D$ mesons and 
$\lambda_4=\frac{1}{3}(e\beta+\frac{2e}{m_c})$  is relevant for loops with charged and strange $D$ mesons.
 Finally, Fig.~\ref{triangle}(e) gives 
\begin{eqnarray}
i\mathcal{M}_{1e} &=& 2g_2^2\lambda_{2(4)}\epsilon_{ijk} q_{i}\epsilon^{\gamma}_{j}\epsilon^{J/\psi}_{k}|\vec{q}\,|^2  \\
&\times& (2I^{(2)}_0(q,m_{D^*},m_{D^*},m_{D^*})+8I^{(2)}_1(q,m_{D^*},m_{D^*},m_{D^*})-I^{(1)}(q,m_{D^*},m_{D^*},m_{D^*})) \, . \nonumber
\end{eqnarray}
In addition to the triangle diagrams with the couplings of $D$ and $D^*$ mesons to the magnetic field from Eq.~(\ref{mag}), there are also two triangle diagrams with the coupling 
of the photon to charged $D$ and $D^*$ mesons that arises due to gauging their kinetic terms. These are shown in  Fig.~\ref{gauge}. The sum of these two diagrams yields 
\begin{eqnarray}
i\mathcal{M}_{2}=4g_2^2e\epsilon_{ijk}q_{i}\epsilon^{\gamma}_{j}\epsilon^{J/\psi}_{k}|\vec{q}\,|^2\left(\frac{1}{m_D}I^{(2)}_1(m_D,m_{D^*},m_D)-\frac{1}{m_{D^*}}I^{(2)}_1(m_{D^*},m_D,m_{D^*})\right).
\end{eqnarray}
So far we have only included the interactions coupling the photon to $D$ and $D^*$ mesons. There are additional diagrams where the photons couple to $\bar{D}$ and $\bar{D}^*$ mesons that  give an equal contribution.

Fig.~\ref{contact} shows the loop diagrams with the contact interaction that arises from gauging the coupling $g_2$.
Fig.~\ref{contact}(a) yields
\begin{eqnarray}
\nonumber
i\mathcal{M}_{3a}&=&i2g_2^2e
\int\frac{d^{4}l}{(2\pi)^4}\frac{\epsilon_{ijk}(2l+q)_{i}\epsilon^{\gamma}_{j}\epsilon^{J/\psi}_{k}}{4(l_{0}-\frac{\vec{l}^2}{2m_D}+i\epsilon)
(l_{0}+q_{0}+\frac{(\vec{l}+\vec{q})^2}{2m_{D^{*}}}+b_{DD^{*}}-i\epsilon)}\\
&=&-g_2^2e\epsilon_{ijk} q_{i}\epsilon^{\gamma}_{j}\epsilon^{J/\psi}_{k}I'(m_D,m_{D^*})\, ,
\end{eqnarray}
where
\begin{eqnarray}
I'(m_D,m_{D^*})=\frac{\mu_{DD^*}}{4\pi}\left(\frac{m_{D^*}-m_D}{{m_{D^*}+m_D}}\right)
\sqrt{\frac{\mu_{DD^*}}{m_D+m_{D^*}}|\vec{q}|^2+2\mu_{DD^*}(b_{DD^*}+q_0)} \, ,
\end{eqnarray}
for $\mu_{DD^*}=m_Dm_{D^*}/(m_D+m_{D^*})$ and $b_{DD^*}=m_D+m_{D^*}-m_{J/\psi}$.
Fig.~\ref{contact}(b) is related to Fig.~\ref{contact}(a) by charge conjugation so 
\begin{eqnarray}
i\mathcal{M}_{3b}=i\mathcal{M}_{3a}=-g_2^2e\epsilon_{ijk} q_{i}\epsilon^{\gamma}_{j}\epsilon^{J/\psi}_{k}I'(m_D,m_{D^*})\, .
\end{eqnarray}
The graphs in Fig.~\ref{contact}(c) and Fig.~\ref{contact}(d)  both vanish, so the total contribution to the  amplitude from loops with contact  interactions is  $2i\mathcal{M}_{3a}$.
Only diagrams with charged and strange $D$ mesons in the loop will contribute.

The total amplitude from loop diagrams in  Fig.~\ref{triangle}  with  neutral $D$ mesons  is given by 
\begin{eqnarray}
\nonumber
i\mathcal{M}^n_{1}&=&\epsilon_{ijk}\vec{q}_{i}\epsilon^{\gamma}_{j}\epsilon^{J/\psi}_{k}\{g_2^2|\vec{q}|^2[4\lambda_1I^{(2)}_1(q,m_D,m_D,m_{D^*})+2\lambda_1(2I^{(2)}_0(q,m_D,m_{D^*},m_{D^*})\\
&&\nonumber+4I^{(2)}_1(q,m_D,m_{D^*},m_{D^*})-I^{(1)}(q,m_D,m_{D^*},m_{D^*}))+2\lambda_1(2I^{(2)}_0(q,m_{D^*},m_{D^*},m_D)\\
&&\nonumber+6I^{(2)}_1(q,m_{D^*},m_{D^*},m_D)-I^{(1)}(q,m_{D^*},m_{D^*},m_D))+2\lambda_2(2I^{(2)}_0(q,m_{D^*},m_D,m_{D^*})\\
&&\nonumber+4I^{(2)}_1(q,m_{D^*},m_D,m_{D^*})-I^{(1)}(q,m_{D^*},m_D,m_{D^*}))+2\lambda_2(2I^{(2)}_0(q,m_{D^*},m_{D^*},m_{D^*})\\
&&+8I^{(2)}_1(q,m_{D^*},m_{D^*},m_{D^*})-I^{(1)}(q,m_{D^*},m_{D^*},m_{D^*}))]\} \, .
\end{eqnarray}
Adding the  five diagrams with the photon coupling to a $\bar{D}$ or $\bar{D}^*$ doubles this contribution. The contribution from diagrams of Fig.~\ref{triangle} with 
 charged and strange charmed mesons in the  loops is obtained by substituting $\lambda_1$ with $\lambda_3$ and $\lambda_2$ with $\lambda_4$.
In addition, the contributions from the diagrams in Fig.~\ref{gauge} and Fig.~\ref{contact} with charged and strange charmed mesons in the loops
 need to be included. The full decay amplitude is
\begin{eqnarray}
i\mathcal{M}_{full}=i\mathcal{M}_0+2(i\mathcal{M}^n_{1}+i\mathcal{M}^c_{1}+i\mathcal{M}^s_{1}+i\mathcal{M}^c_{2}+i\mathcal{M}^s_{2}+i\mathcal{M}^c_{3a}+i\mathcal{M}^s_{3a}) \, ,
\end{eqnarray}
where the superscript $c (s)$ indicates a contribution from loops with charged (strange) $D$ mesons.
The decay rate for $J/\psi\rightarrow\gamma\eta_c$ is given by
\begin{eqnarray}
\Gamma[J/\psi\rightarrow\gamma\eta_c]=\frac{1}{8\pi}(\sqrt{m_{J/\psi}m_{\eta_c}}M_{full})^2\frac{|\vec{q}\,|}{m_{J/\psi}^2}.
\end{eqnarray}


\end{document}